\documentclass[english]{article}
\usepackage{amsmath, amsfonts, amssymb}
\usepackage[utf8]{inputenc}
\usepackage{graphicx}
\usepackage{geometry}
\usepackage[round]{natbib}
\usepackage{subfig} 
\usepackage{hyperref} 
\usepackage{physics}
\usepackage{float}
\usepackage{authblk}
\usepackage{caption}
\usepackage{cleveref}

\hypersetup{
	bookmarksopen=true,
	colorlinks=true,
	linkcolor=blue,
	citecolor=blue,
	urlcolor=black,	
	pdftitle={Waves in C-CST media},
	pdfauthor={N. Guarin-Zapata, J. Gomez, A.R. Hadjesfandiari, G. Dargush},
	pdfkeywords={Finite element method, Couple-stress, Bloch theorem, Wave 
	propagation},
	pdfsubject={Elastodynamics},
	pdfpagemode=UseOutlines,
	pdfstartview=FitH
}

\title{\textbf{Variational principles and finite element Bloch analysis in
couple stress elastodynamics}}

\author[1]{Nicolás Guarín-Zapata \thanks{nguarinz@eafit.edu.co}}
\author[1]{Juan Gomez \thanks{jgomezc1@eafit.edu.co}}
\author[2]{Ali Reza Hadjesfandiari \thanks{a.hadjesfandiari@ccsu.edu}}
\author[3]{Gary F. Dargush \thanks{gdargush@buffalo.edu}}
\affil[1]{Departamento de Ingeniería Civil, Universidad EAFIT}
\affil[2]{Department of Engineering, Central Connecticut State University}
\affil[3]{Department of Mechanical and Aerospace Engineering, University at Buffalo}

\begin{document}
\maketitle

\abstract{We address the numerical simulation of periodic solids (phononic
crystals) within the framework of couple stress elasticity. The additional
terms in the elastic potential energy lead to dispersive behavior in shear
waves, even in the absence of material periodicity. To study the bulk waves in
these materials, we establish an action principle in the frequency domain and
present a finite element formulation for the wave propagation problem related
to couple stress theory subject to an extended set of Bloch-periodic boundary
conditions. A major difference from the traditional finite element formulation
for phononic crystals is the appearance of higher-order derivatives. We solve
this problem with the use of a Lagrange-multiplier approach. After presenting
the variational principle and general finite element treatment, we particularize
it to the problem of finding dispersion relations in elastic bodies with
periodic material properties. The resulting implementation is used to determine
the dispersion curves for homogeneous and porous couple stress solids, in which
the latter is found to exhibit an interesting bandgap structure.}

\textbf{Keywords:} Micromechanics, phononic crystals, couple stress elasticity,
wave propagation, dispersive media, metamaterials.

\setlength{\parindent}{0 mm}
\setlength{\parskip}{14 pt}

\pagebreak
\section{Introduction}

There has been significant interest, especially in recent years, to develop
spatially periodic band gap materials and structures, based upon Floquet-Bloch
theory \citep{floquet1883, bloch1929}. Recent developments in the field of
architectured materials aimed at achieving novel mechanical properties often
rely on enhancements that include effects neglected by classical theories.
Continuum models with local microstructural interactions have become
increasingly popular after the advance and growth in the field of metamaterials,
as summarized in the monograph by  \cite{book:banerjee_metamaterials}. A family
of models that has regained popularity in the last few years is the so-called
Cosserat-based theories, which are mainly founded on the formulation by
\cite{Cosserat1909}. In a wide sense, these material models consider
microstructural effects through a generalization of Cauchy's postulate to
include additional mechanical interactions involving couples per unit surface
or couple-stresses. In the present work, we focus on a pure continuum mechanics
representation by widening the modeling capabilities of consistent-couple stress
theory (C-CST), originally formulated in \cite{hadjesfandiari2011couple}. In
particular, we establish for the first time a principle of stationary correlated
action for the corresponding reduced wave equation of elastodynamics and extend
the theory to spatially periodic materials, thus providing an objective physical
basis to characterize material through its dispersive behaviour.

The entire family of Cosserat elasticity models depart from the classical Cauchy
models in the consideration of microstructural effects, which are unavoidably
expected to occur once the specimen dimensions become comparable to the material
microstructural features. These effects cannot be addressed in classical
theories. On the other hand, microstructural effects are introduced through the
extension of dynamic and kinematic descriptors from classical continuum mechanics
on a range of alternative models. \cite{Voigt1910} was probably the first to
postulate a model with asymmetric mechanical interactions in terms of
couple-stresses: the interaction between two material points in this continuum
encompassed couples per unit contact surface in addition to the classical Cauchy
forces per unit surface. In a landmark contribution, the Cosserat brothers
\citep{Cosserat1909} formulated a mathematical theory involving couple-stresses
in which new kinematic variables were introduced in the form independent
micro-rotations. Various later extensions from these theories were also
developed by \cite{Eringen1966, book:nowacki1986, Mindlin1964,
eringen1964nonlinear} in micropolar, microstretch and micromorphic theories. 
Alongside, a different branch of developments resulted in a set of couple stress
theories in the work by \cite{Toupin1962}, \cite{MindlinAndTiersten1962} and
\cite{Koiter1964}, who used the gradients of the true continuum rotation field
to provide the required kinematic enrichment.

Developments from \cite{hadjesfandiari2011couple} have resulted in a
consistent version of the models by \cite{Toupin1962},
\cite{MindlinAndTiersten1962} and \cite{Koiter1964} in terms of couple-stresses.
The consistency of this model is reflected in the determinacy of all the
force-stress and couple-stress components, the identification of the necessary
and sufficient set of natural and essential boundary conditions and the
elimination of redundant force components. An approach for evaluating the
usefulness and robustness of a continuum mechanics model is through the
determination of its band structure in terms of its dispersion relationships.
These indicate the kinematic response of the material through an identification
of the wave propagation modes that can exist within the model and the frequency
dependency of the group and phase velocities of these potential waves. An
effective technique, relying on the assumption of spatial periodicity, is based
on Bloch's theorem from solid state physics \citep{book:brillouin}, where the
problem of finding the band structure reduces to solving a series of generalized
eigenvalue problems for a variation of the wave vector in the reciprocal space.
In the case of the C-CST model, this problem poses several computational
challenges. First, since the enriched kinematic variables are now curvatures,
corresponding to particular second order gradients of the displacement field,
the displacement-based finite element formulation now would require \(C^1\)
interelement continuity. As shown by \cite{darrall2014}, this numerical issue
can be resolved by introducing Lagrange multiplier techniques, however it is
not obvious how to incorporate these within Bloch analysis. Second, as a result
of enforcing the kinematic constraint in terms of Lagrange multipliers, the
computational framework lacks inertial components associated with the rotational
interactions. Since there is only a mass matrix associated with the
translational degrees of freedom, special attention is needed in solving the
eigenproblem. Both of these issues are resolved in the present
work.

The characterization of the bulk properties of periodic materials is commonly
done finding the band structure or dispersion relations \citep{hussein2014dynamics}.
Commonly, this band structure is obtained using a numerical method such as 
the Boundary Element Method \citep{li2013bandgap, li2013boundary}, the Finite
Difference Method \citep{tanaka2000band, su2010postprocessing,
isakari2016periodic}, the Finite Element Method \citep{langlet95, guarin2015, 
valencia_uel_2019, mazzotti2019modeling, guarin2020,chin2021spectral},  or
the Plane Wave Expansions \citep{cao2004convergence, xie2017improved,
dal2020elastic}. We favor the use of the Finite Element Method because of its
maturity and versatily to represent arbitrary geometries and boundary
conditions. In this work, we find the dispersion relations modeling
a single unit cell of the material and using Bloch's theorem. There have been
few works on periodic materials involving generalized continua and these have
been related to micropolar elasticity \citep{zhang2018, guarin2020}. To the best
of our knowledge, this is the first work using a higher-order elasticity model
for phononic crystals.

Here we establish a new variational principle in the temporal frequency domain
for reduced couple stress elastodynamics and then extend the finite element
algorithm from \cite{darrall2014} to the case of spatially periodic material
cells with Bloch boundary conditions. We examine first the closed form
dispersion relationships for the homogeneous version of the model. This
homogeneous model already involves micromechanical effects through a length
scale material parameter, however additional effects can be considered in terms
of explicit representations of geometric features at the fundamental material
cell level. We then formulate a variational statement together with the
imposition of an extended version of the usual Bloch periodic boundary
conditions that satisfies Hermiticity and positive definiteness for C-CST.
Subsequently, this statement is modified by introducing an artificial
independent rotation field tied to the continuum displacement field through the
enforcement of a Lagrange multiplier field that is shown to equal the
skew-symmetric part of the force-stresses. The resulting numerical framework is
tested by comparing its results with those obtained in closed form
for the homogeneous case and by applying it to a porous periodic material cell
design, which displays interesting bandgap behavior that has
not been resported previously.

\section{Governing equations}

\subsection{Forces and moments in the C-CST solid}
\begin{figure}[H]
\centering
\includegraphics[width=5 cm]{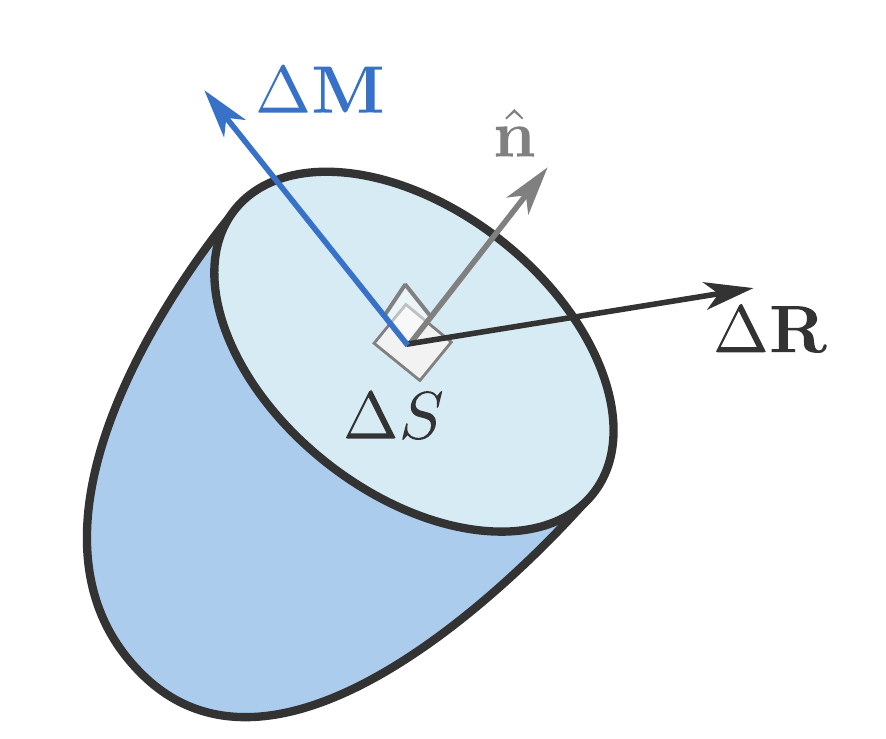}
\caption{In the C-CST model mechanical effects are described through linear and
rotational interactions in terms of resultant forces and moments at the material
point. These resultants act on the surface element \(\Delta S\) over a plane
whose unit outward normal is \(\hat{n}\).}
\label{fig:cauchy_principle}
\end{figure}

The fundamental signature of the extended continuum model considered in this
work is the presence of rotational mechanical interaction, in addition to the
classical translational interaction between material points in the continuum.
Following a generalized Cauchy's postulate \citep{MindlinAndTiersten1962,
Koiter1964} we define force and couple traction vectors \(t_i^{(\hat{n})}\)
and \(m_i^{(\hat{n})}\) respectively as
\begin{subequations}
\label{eq:tracts}
\begin{align}
        &t_i^{(\hat{n})} = \lim_{\Delta S(\hat{n}) \rightarrow 0} \frac{\Delta 
        R_i}{\Delta S(\hat{n})}\label{eq:for_T} \\
        &m_i^{(\hat{n})} = \lim_{\Delta S(\hat{n}) \rightarrow 0} \frac{\Delta 
        M_i}{\Delta S(\hat{n})} \, , \label{eq:for_M}       
\end{align}
\end{subequations}
and where \(\Delta S(\hat{n})\) is a small element of area oriented with unit
normal \(\hat{n}\) while \(\Delta R_i\) and \(\Delta M_i\) are the resultant
force and couple moment, respectively. However, only the tangential components of
\(m_i^{(\hat{n})}\) exist as independent bending couple tractions
(\cref{fig:cauchy_principle}).

Note that while the force-tractions vector \(t_i^{(\hat{n})}\) is a polar vector,
the couple-tractions vector \(m_i^{(\hat{n})}\) is an axial vector.
Force-tractions and couple-tractions are also described by projections of the
non-symmetric force-stress tensor \(\sigma_{ij}\) and the couple-stress
\(\mu_{ij}\) tensors according to:
\begin{subequations}
\label{eq:tensions}
\begin{align}
        &t_i^{(\hat{n})} = \sigma_{ji} n_j\, ,\label{eq:for_sigma}\\
        &m_i^{(\hat{n})} = \mu_{ji} n_j=\epsilon_{ijk}\mu_k n_j\, , \label{eq:cou_sigma}       
\end{align}
\end{subequations}
where \(\mu_{ij}\) is skew-symmetric. Thus, \(\mu_{ij} = -\mu_{ji}\) and the
couple-stress tensor can be written as a polar vector with
\[\mu_k = \frac{1}{2}\epsilon_{kji}\mu_{ji}\, ,\]
where \(\epsilon_{ijk}\) is the Levi-Civita permutation symbol leading to the
last form in \eqref{eq:cou_sigma}, which clearly shows that \(m_i^{(\hat{n})}\)
is tangential to the surface.

Consideration of the linear and angular balance equations for an arbitrary part
of the material continuum of volume \(V\), bounded by external surface \(S\)
leads to the following force-stress and couple-stress equilibrium equations for
the C-CST model:
\begin{equation}
\begin{split}
&\sigma_{ji,j} + f_i = \rho \ddot{u}_i\, ,\\
&\mu_{ji,j} + \epsilon_{ijk}\sigma_{jk} = 0\, ,
\end{split}
\label{eq:balance_eqs}
\end{equation}
where \(f_i\) are forces per unit volume, and \(\rho\) is the mass density.
Notice that, in contrast to micropolar models \citep{guarin2020} where there is
a rotational inertial density and a body couple term, in this model the balance
equations already include those contributions. This particular aspect of the
C-CST model is discussed in the original paper by \cite{hadjesfandiari2011couple}
where it is also proved that from
\begin{equation}
\epsilon_{ijk}(\mu_{k,j} + \sigma_{jk}) = 0\, ,
\label{eq:skew_symmetry}
\end{equation}
it follows that \(\mu_{k,j} + \sigma_{jk}\) is symmetric and as a result its
skew-symmetric part is zero leading to
\[\sigma_{[ji]} = - \mu_{[i,j]}\, .\]
This gives the skew-symmetric part of the force-stress tensor in terms of the
couple-stress vector, which also can be described by its dual vector
representation
\[s_i = \frac{1}{2} \epsilon_{ijk} \mu_{k, j}\, .\]

\subsection{Kinematics and constitutive relations}
In the linear C-CST model, kinematics is described by the classical infinitesimal
strain (\(e_{ij}\)) and rotation (\(\theta_{ij}\)) tensors
\begin{subequations}
\label{eq:kinematics}
\begin{align}
e_{ij} = \frac{1}{2}(u_{i,j} + u_{j,i})\, , \label{eq:strain}\\
\theta_{ij} = \frac{1}{2}(u_{i,j} - u_{j,i})\, , \label{eq:rotation}
\end{align}
\end{subequations}
and by the mean curvature tensor
\begin{equation}
\kappa_{ij} = \frac{1}{2}(\theta_{i,j} - \theta_{j,i})\, ,
\label{eq:curv_tensor}
\end{equation}
where
\[\theta_i = \frac{1}{2}\epsilon_{ijk}\theta_{kj}\, .\]

Equation \eqref{eq:curv_tensor} can also be written in polar form, as an
engineering curvature vector \citep{darrall2014}
\begin{equation}
\kappa_i = \epsilon_{ijk} \theta_{j,k} =\frac{1}{2}(u_{i,kk} - 
u_{k,ik})
\label{eq:curv}
\end{equation}
since
\[\kappa_i = \epsilon_{ijk}\kappa_{jk}\, .\]

For a linear elastic centrosymmetric C-CST continuum, the constitutive equations
can be written as
\begin{equation}
\begin{split}
&\sigma_{(ij)} = C_{ijkl} e_{kl}\, ,\\
&\mu_i = D_{ij} \kappa_j\, ,
\end{split}
\label{eq:constitutive_aniso}
\end{equation}
where \(C_{ijjkl}\) is the stiffness tensor as in classical (anisotropic)
elasticity, and \(D_{ij}\) is an additional material tensor that accounts for 
couple-stress effects. In the expressions above, parentheses as subindices are
used to indicate the symmetric part of the tensor. In the case of a linear
isotropic elastic C-CST continuum,
\begin{equation}
\begin{split}
&C_{ijkl} = \lambda \delta_{ij} \delta_{kl} + \mu ( \delta_{ik} \delta_{jl}
   + \delta_{il} \delta_{jk} ), \\
&D_{ij} = 4\eta \delta_{ij},
\end{split}
\label{eq:isotropic_tensors}
\end{equation}
where \(\mu\) and \(\lambda\) are the Lamé parameters as in classical 
elasticity, while \(\eta\) is the additional material coefficient that accounts
for couple-stress effects. Then, the constitutive equations for isotropy can be
simplified to
\begin{equation}
\begin{split}
&\sigma_{(ij)} = \lambda e_{kk} \delta_{ij} + 2\mu e_{ij}\, ,\\
&\mu_i = 4\eta \kappa_i\, .
\end{split}
\label{eq:constitutive}
\end{equation}

\subsection{Displacement equations of motion}
At this point it may be convenient to alternate between index and explicit
vector notation. In the latter, the gradient operator reads
\(\nabla = \frac\partial{\partial x_i}\) in Cartesian coordinates. In these
terms, the time domain displacement equations of motion are obtained after
using the constitutive relations \eqref{eq:constitutive} in the equilibrium
equations \eqref{eq:balance_eqs} yielding
\begin{equation}
(\lambda + 2\mu)\nabla (\nabla \cdot \mathbf{u}) - \mu\nabla\times \nabla 
\times \mathbf{u} + \eta \nabla^2 \nabla \times \nabla \times \mathbf{u} = 
\rho\ddot{\mathbf{u}}\, .
\label{eq:wave_eq}
\end{equation}

Defining the phase/group speed for the longitudinal (P) wave \(c_1\) (which is
not dispersive), the low-frequency (\(k\rightarrow 0\)) phase/group speed for
the transverse wave (S) \(c_2\) (which is dispersive) and the intrinsic material
length scale parameter \(l\) (which is not present in classical elasticity),
such that
\begin{equation}
c_1^2 = \frac{\lambda + 2\mu}{\rho}\, ,\qquad
c_2^2 = \frac{\mu}{\rho}\, ,\qquad
l^2 = \frac{\eta}{\mu}\, ,
\label{eq:c1c2l2}
\end{equation}
allows us to write \eqref{eq:wave_eq} in the form
\begin{equation}
c_1^2 \nabla (\nabla \cdot \mathbf{u}) - c_2^2 (1-l^2 \nabla^2) \nabla\times \nabla \times 
\mathbf{u} = \ddot{\mathbf{u}}\, .
\label{eq:wave_eq_speeds}
\end{equation}

\subsection{Dispersion relations for unbounded domains}
Using a Helmholtz decomposition,  the displacement field can be written in terms
of the scalar and vector potentials \(\varphi\) and \(\mathbf{H}\)
\citep{book:arfken} as
\[\mathbf{u} = \nabla\varphi + \nabla\times\mathbf{H}\, ,\quad 
\nabla\cdot\mathbf{H} = 0\, ,\]
and replacing this in \eqref{eq:wave_eq_speeds} gives the following set of
uncoupled wave equations
\begin{align}
&c_1^2\nabla^2\varphi = \ddot{\varphi}\, ,\\
&c_2^2(1 - l^2\nabla^2)\mathbf{H} = \ddot{\mathbf{H}}\, ,
\end{align}
where it is observed that the equation for the rotational potential follows a
higher-order wave equation that is inherently dispersive.  This becomes evident
after assuming a solution of the form \(\mathbf{u} = \mathbf{\tilde u}e^{ik x - 
i\omega t}\) which gives the dispersion relations
\begin{align}\label{eq:dispersion}
&\omega_P^2 = c_1^2 k^2\, ,\\
&\omega_S^2 = c_2^2 k^2 (1 + k^2 l^2)\, . \label{eq:dispersion_SV}
\end{align}

Solving the above for \(k\), we have in each case
\begin{align*}
k_P^2 &= \frac{\omega^2}{c_1^2}\, , &k_S^2 &= \frac{1}{2 l^2}\left[\pm\sqrt{1 
+ \frac{4\omega^2 l^2}{c_2^2}} - 
1\right]\, .
\end{align*}

Noticing that the quantity inside the square root is always greater than 1
indicates that we should consider only the positive root, while the negative
root corresponds to an evanescent wave that should arise under certain boundary
conditions. The phase and group speeds are now given by
\begin{equation}
\begin{aligned}
v_P &= c_1\, , & g_P &= c_1\, ,\\
v_S(k) &= c_2\sqrt{1 + k^2 l^2}\, , & g_S &= c_2 \frac{1 + 2k^2 l^2}{\sqrt{1 + 
k^2 l^2}}\, .
\end{aligned}
\label{eq:speeds}
\end{equation}

\begin{figure}[H]
    \centering
    \includegraphics[height=1.8 in]{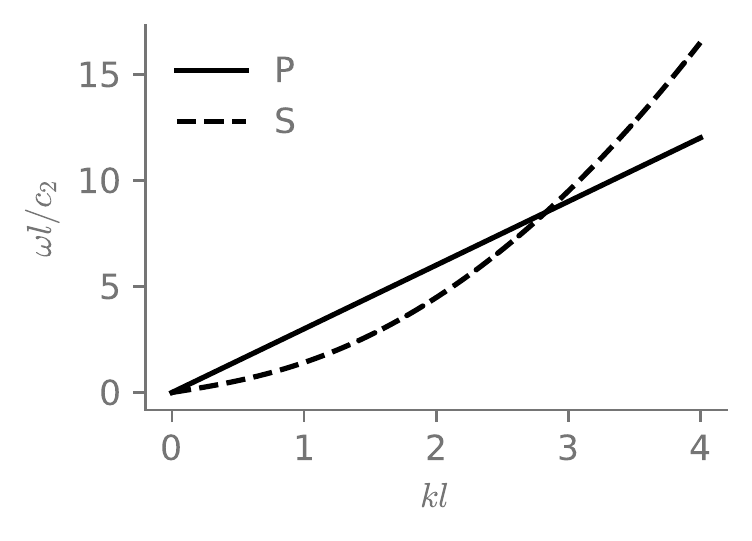}
    \includegraphics[height=1.8 in]{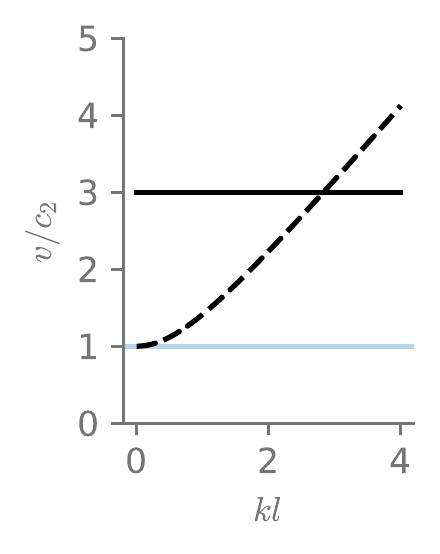}
    \includegraphics[height=1.8 in]{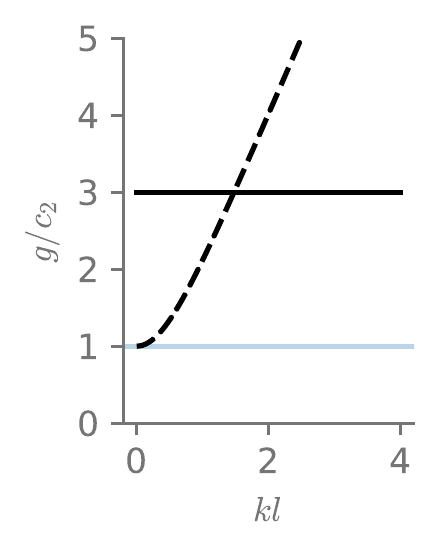}
    \caption{Dispersion relations for a homogeneous C-CST material 
        with properties: \(\rho = 1\times 10^5\), \(\lambda=2.8 \times 10^{10}\),
        \(\eta=1.62 \times 10^9\), \(\mu = 4\times 10^9\). The plot  on 
        the left shows the frequency-wave number relation for the 
        non-dispersive P-wave (continuous line) and the dispersive SV (dashed
        lines). The plots in the middle and right part of the  figure show the
        phase and group speeds for the dispersive modes.}
    \label{fig:dispersion_analytic}
\end{figure}

Taking the low and high frequency limits \(k \rightarrow 0\) and
\(k \rightarrow \infty\) gives
\begin{align*}
&\lim_{k \rightarrow 0} v_S = \lim_{k \rightarrow 0} g_S = 
c_2\, ,\\
&\lim_{k \rightarrow \infty} v_S = \lim_{k \rightarrow \infty}g_S 
\rightarrow  \infty\, ,
\end{align*}
which shows how the speed of energy flow increases with frequency. All of these 
relations are displayed in \cref{fig:dispersion_analytic}.

\subsection{Frequency domain equations}
Bloch analysis considering spatial periodicity of the material is naturally
conducted in the Fourier domain, involving both the temporal frequencies and
spatial wave numbers.  After performing a Fourier transform of the linear and
angular momentum equations  \eqref{eq:balance_eqs} to the temporal frequency
domain, these become
\begin{equation}
\begin{split}
&\tilde\sigma_{ji,j} + \tilde f_i = - \rho \omega^2 \tilde{u}_i\, ,\\
&\tilde\mu_{ji,j} + \epsilon_{ijk}\tilde\sigma_{jk} = 0\, ,
\end{split}
\label{eq:fourier_balance_eqs}
\end{equation}
where the superposed tilde denotes a complex Fourier amplitude.

After introducing the constitutive equations \eqref{eq:constitutive_aniso} for
centrosymmetric materials into \eqref{eq:fourier_balance_eqs} and then combining
the angular momentum and linear momentum balance laws into a single set in terms
of displacement, one finds:
\begin{equation}
( C_{ijkl} \tilde u_{k,l} )_{,j} + \frac{1}{4}\epsilon_{pij}\epsilon_{pmn}\{D_{nk} (\tilde u_{k,ll}-\tilde u_{l,kl})\}_{,mj} + f_i = -\rho \omega^2 \tilde u_i
\label{eq:fourier_disp_aniso}
\end{equation}

Substituting \eqref{eq:isotropic_tensors} and \eqref{eq:c1c2l2} for isotropic
materials into \eqref{eq:fourier_disp_aniso} provides the corresponding Fourier
domain reduced wave equations in the absence of body forces, which can be written
\begin{equation}
c_1^2 \nabla (\nabla \cdot \mathbf{\tilde u}) - c_2^2 (1-l^2 \nabla^2) \nabla\times \nabla \times
\mathbf{\tilde u}  = -\omega^2 \mathbf{\tilde u}\, .
\label{eq:wave_eq_freq}
\end{equation}
Notice that \eqref{eq:wave_eq_freq} is the temporal Fourier transform of
\eqref{eq:wave_eq_speeds}.

\section{Variational principles}
We will describe next a variational formulation for the elastodynamic C-CST
model. An inherent complexity is the presence of second order displacement
gradients arising in the curvatures \eqref{eq:curv}, which requires \(C^1\)
continuity of the displacement field.

\begin{figure}[H]
    \centering
    \includegraphics[width=4 in]{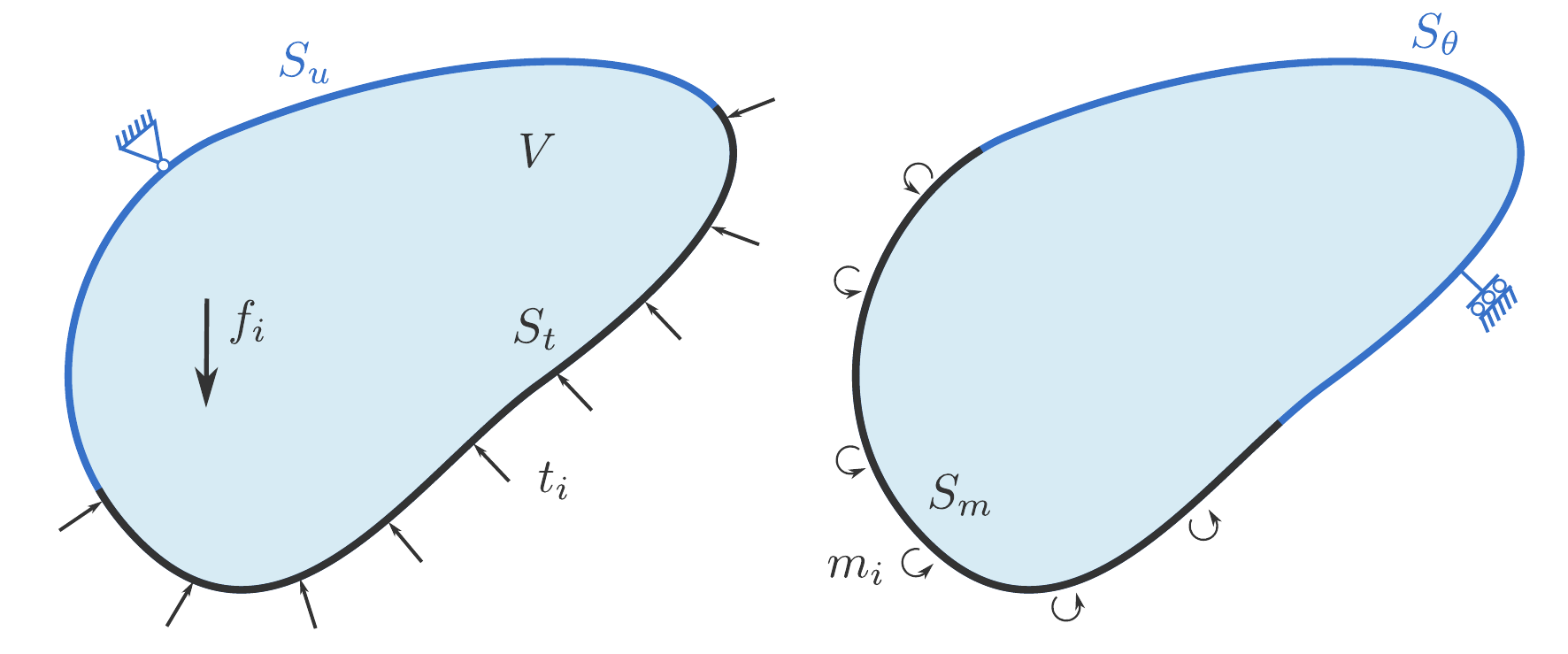}
    \caption{Schematic representation of the domain and boundary conditions for the C-CST model.}
    \label{fig:domain}
\end{figure}

Let us consider a volume \(V\) with boundary \(S\), having specified body forces
\(f_i\), force-tractions \(t_i\), and couple-tractions \(m_i\) (see
\cref{fig:domain}). The boundary \(S\) is split into different segments, where
\(S_u\) represents the portion of \(S\) with specified displacements, \(S_t\)
represents the surface with prescribed tractions, \(S_\theta\) represents the
segment with enforced rotations, and \(S_m\) the boundary with prescribed
couple-tractions. Additionally, \(S = S_u \cup S_t = S_\theta \cup S_m\) and
\(S_u \cap S_t = S_\theta \cap S_m = \emptyset\). In general, \(S_u\) and \(S_t\)
might overlap with \(S_\theta\) and \(S_m\). This is an important aspect of the
C-CST model that is relevant in the solution of boundary value problems, as we
shall see later.

We begin with the following couple stress elastodynamic action functional in the
frequency domain, as an extension of the elastostatic formulations introduced
by \cite{hadjesfandiari2011couple} and \cite{darrall2014}:
\begin{equation}
\mathcal{A}[u;\omega] = \mathcal{U}[u;\omega]+ \mathcal{T}[u;\omega]+\mathcal{V}[u;\omega],
\label{eq:Lagrangian_functional}
\end{equation}
Here, and in the remainder of this paper, the superposed tilde has been
suppressed for notational convenience.  Meanwhile, the elastic, kinetic and
applied load actions can be written in explicit form, respectively, as
\begin{equation}
\mathcal{U}[u;\omega] = \frac{1}{2}\int\limits_{V} e_{ij}^* C_{ijkl} e_{kl} \dd{V}
+ \frac{1}{2}\int\limits_{V} \kappa_{i}^* D_{ij} \kappa_{j} \dd{V},
\label{eq:delta_functional_U}
\end{equation}
\begin{equation}
\mathcal{T}[u;\omega] = -\frac{\omega^2}{2} \int\limits_{V} u_i^* \rho u_i \dd{V},
\label{eq:delta_functional_T}
\end{equation}
\begin{equation}
\mathcal{V}[u;\omega] =- \int\limits_{V} u_i^* f_i \dd{V}
- \int\limits_{S_t}  u_i^* t_i^{(\hat{n})} \dd{S}  - \int\limits_{S_m} \theta_i^* m_i^{(\hat{n})} \dd{S}. 
\label{eq:delta_functional_V}
\end{equation}
with the asterisk denoting complex conjugate.

The stationarity of this action becomes
\begin{equation}
\delta \mathcal{A}[u;\omega] = \delta\mathcal{U}[u;\omega]
+ \delta\mathcal{T}[u;\omega]+\delta\mathcal{V}[u,\omega]) = 0,
\label{eq:action_variation}
\end{equation}
or
\begin{equation}
\begin{split}
&\delta\mathcal{A}[u;\omega] = 
 \int\limits_{V} \delta e_{ij}^* C_{ijkl} e_{kl} \dd{V}
 + \int\limits_{V} \delta\kappa_{i}^* D_{ij} \kappa_{j} \dd{V}
 -\omega^2 \int\limits_{V} \delta u_i^* \rho u_i \dd{V} \\
&- \int\limits_{V} \delta u_i^* f_i \dd{V}
- \int\limits_{S_t}  \delta u_i^* t_i^{(\hat{n})} \dd{S}
- \int\limits_{S_m} \delta\theta_i^* m_i^{(\hat{n})} \dd{S} = 0, \,
\label{eq:delta_action_functional}
\end{split}
\end{equation}
which can serve as the weak form for a finite element formulation in reduced
elastodynamics.  With the appearance of mean curvature in
\eqref{eq:delta_action_functional}, this would require \(C^1\) spatial continuity
of displacements \(u_i\).

Next, let us derive the Euler-Lagrange equations associated with the functional
\(\mathcal{A}[u; \omega]\). Starting from the first variation in
\eqref{eq:delta_action_functional}, we repeatedly apply integration-by-parts
operations and the divergence theorem to shift all of the spatial derivatives
from the variations to the true fields. This leads to the following
statement:
\begin{equation}
\begin{split}
&\int\limits_{V} \delta u_i^* [ ( C_{ijkl} u_{k,l} )_{,j}
  + \frac{1}{4}\epsilon_{pij}\epsilon_{pmn}\{D_{nk} (u_{k,ll}-u_{l,kl})\}_{,mj}
  + f_i + \rho \omega^2 u_i] \dd{V} \\
&+ \int\limits_{S_t}  \delta u_i^* [ t_i^{(\hat{n})} - \sigma_{ji} n_j ] \dd{S} \\
&+ \int\limits_{S_m} \delta\theta_i^* [ m_i^{(\hat{n})} -  \epsilon_{ijk}\mu_k n_j ] \dd{S} = 0. \, 
\label{eq:delta_action_E-L_functional}
\end{split}
\end{equation}
For arbitrary variations, each set of terms inside the square brackets must be
zero. Thus, the Euler-Lagrange equations can be written:
\begin{equation}
( C_{ijkl} u_{k,l} )_{,j} + \frac{1}{4}\epsilon_{pij}\epsilon_{pmn}\{D_{nk} (u_{k,ll}-u_{l,kl})\}_{,mj}
  + f_i = -\rho \omega^2 u_i \quad {\rm in} \ V \\
\label{eq:action_E-L_aniso_momentum}
\end{equation}
\begin{equation}
\begin{split}
&t_i^{(\hat{n})} = \sigma_{ji} n_j \quad {\rm on} \ S_t \\
&m_i^{(\hat{n})} = \epsilon_{ijk}\mu_k n_j \quad {\rm on} \ S_m \,
\label{eq:action_E-L_aniso_traction}
\end{split}
\end{equation}
Notice that \eqref{eq:action_E-L_aniso_momentum} are the reduced wave equations
from \eqref{eq:fourier_disp_aniso}, while \eqref{eq:action_E-L_aniso_traction}
represent the corresponding natural boundary conditions for C-CST.  In the
isotropic case, substituting \eqref{eq:isotropic_tensors} into
\eqref{eq:action_E-L_aniso_momentum} produces
\begin{equation}
c_1^2 u_{j,ji}-c_2^2 \epsilon_{ijk} \epsilon_{kmn} (u_{n,mj}-l^2 u_{n,mjll})
+ f_i = -\omega^2 u_i\quad {\rm in} \ V
\label{eq:action_E-L_iso_momentum}
\end{equation}
which is the equivalent of \eqref{eq:wave_eq_freq} in index notation.

Performing an inverse Fourier transform of individual terms in
\eqref{eq:Lagrangian_functional}-\eqref{eq:delta_functional_V} back to the time domain, one finds
\begin{equation}
\mathcal{F}^{-1}[u^* v] = (u \star v)(t)
\label{eq:time_domain_functional}
\end{equation}
where the $\star$ operator denotes correlation over time, such that
\begin{equation}
(u \star v)(t) = \int_{-\infty}^\infty u(\tau) v(t+\tau) d\tau.
\label{eq:correlation}
\end{equation}

Consequently, we have established the following stationary {\it Principle of
Correlated Action} for couple stress elastodynamics: Of all the possible
displacement fields in \(V\) that satisfy the frequency domain kinematic boundary
conditions on \(S_u\) and \(S_\theta\), the one that renders the action
\(\mathcal{A}[u; \omega]\) in \eqref{eq:Lagrangian_functional} stationary
corresponds to the solution of the reduced wave equations
\eqref{eq:action_E-L_aniso_momentum} and traction boundary conditions
\eqref{eq:action_E-L_aniso_traction}.

We should emphasize that this stationary Principle of Correlated Action also
holds for classical theory, if one neglects contributions from mean curvature
and moment tractions.  Thus, the classical correlated action for reduced
elastodynamics can be written:
\begin{equation}
\begin{split}
&\mathcal{A}_{\rm cl}[u;\omega] = \frac{1}{2}
 \int\limits_{V} e_{ij}^* C_{ijkl} e_{kl} \dd{V}
 -\frac{\omega^2}{2} \int\limits_{V} u_i^* \rho u_i \dd{V} \\
&\qquad\qquad- \int\limits_{V} u_i^* f_i \dd{V}
- \int\limits_{S_t} u_i^* t_i^{(\hat{n})} \dd{S} . \,
\label{eq:classical_action_functional}
\end{split}
\end{equation}

\section{Response of Periodic Materials}
This section summarizes the most relevant theoretical aspects for the numerical
analysis of periodic materials. An in-depth treatment of the subject can be
found in classical textbooks, such as \cite{book:brillouin} and \cite{book:kittel},
while a comprehensive review is provided in \cite{hussein2014dynamics}. In our
discussion we will use a generalized form of the reduced wave equation, however
we will provide the Bloch-Floquet boundary conditions for the particular case of
the C-CST model.

\subsection{Bloch's theorem}
Consider a reduced elastodynamic wave equation in the frequency domain of the form
\begin{equation}
  \mathcal{L} \mathbf{u}(\mathbf{x}) = -\rho \omega^2 \mathbf{u}(\mathbf{x})\ 
  \label{eq:reduced}
\end{equation}
valid for a  field \(\mathbf{u}\) at a spatial point \(\mathbf{x}\). Here
\(\mathcal{L}\) is a positive definite linear differential operator
\citep{book:reddy_functional, book:kreyszig_functional, johnson2007waves},
while \(\rho\) is the mass density and \(\omega\) the corresponding angular
frequency. Bloch's theorem \citep{book:brillouin} establishes that solutions to
\eqref{eq:reduced} are of the form
\begin{equation}
  \mathbf{u}(\mathbf{x}) = \mathbf{w}(\mathbf{x}) e^{i\mathbf{k}\cdot \mathbf{x}}\
  \label{eq:bloch}
\end{equation}
where $\mathbf{w}(\mathbf{x})$ is a Bloch function carrying with it the same
periodicity as the material. Since the spatial period in
\(\mathbf{w}(\mathbf{x})\) is the lattice parameter \(\mathbf{a}\), it follows that
\[\mathbf{w}(\mathbf{x} + \mathbf{a}) = \mathbf{w}(\mathbf{x}).\]

Accordingly, \eqref{eq:bloch} is the product of a spatially periodic function
\(\mathbf{w}(\mathbf{x})\), with the periodicity of the lattice, and a plane
wave (of wave vector \(\textbf{k}\)), which is also periodic. As a result, field
variables \(\mathbf{\Phi}\) at opposite sides of the unit cell and separated by
the lattice vector \(\mathbf{a}\) are related through
\begin{equation}
\mathbf{\Phi}(\mathbf{x} + \mathbf{a}) = \mathbf{\Phi}(\mathbf{x})e^{i\mathbf{k}\cdot\mathbf{a}} .
\label{eq:gen_bloch}
\end{equation}

In this case, $\mathbf{\Phi}$ refers to the principal variable involved in the
physical problem, or to any of its spatial derivatives. From a physical point
of view, \eqref{eq:gen_bloch} means that a field variable \(\mathbf{\Phi}\) at
points \(\mathbf{x}\) and \(\mathbf{x} + \mathbf{a}\) differ only by the phase
shift \(e^{i\mathbf{k}\cdot\mathbf{a}}\).

In the classical elastodynamic case in which \(\mathcal{L}\) is the Navier
operator of order 2, the generalized Boundary Value Problem (BVP) considering
Bloch boundary conditions (BBCs) takes the form:
\begin{subequations}
\label{eq:bvp_bloch}
\begin{align}
&\mathcal{L} \mathbf{u}(\mathbf{x}) = -\rho \omega^2 \mathbf{u}(\mathbf{x})\, ,\\
&\mathbf{u}(\mathbf{x} + \mathbf{a}) = \mathbf{u}(\mathbf{x})e^{i\mathbf{k}\cdot\mathbf{a}}\, ,\\
&\mathbf{\sigma}(\mathbf{x} + \mathbf{a}) \cdot \hat{\mathbf{n}}
  = - \mathbf{\sigma}(\mathbf{x}) \cdot \hat{\mathbf{n}}\,  e^{i\mathbf{k}\cdot\mathbf{a}}\, ,
\end{align}
\end{subequations}
where \(\mathbf{u}(\mathbf{x} + \mathbf{a})\) and \(\mathbf{u}(\mathbf{x})\)
give the field at \(\mathbf{x} + \mathbf{a}\) and \(\mathbf{x}\), respectively,
and \(\mathbf{\sigma}(\mathbf{x})\) is the corresponding stress.  Meanwhile, 
\(\mathbf{a} = \mathbf{a}_1 n_1 + \mathbf{a}_2 n_2 + \mathbf{a}_3 n_3\) is the
lattice translation vector and \(n_i\) are the lattice normal parameters.

Note that the BVP encompassed by \eqref{eq:bvp_bloch} simultaneously describes
the space-time periodicity of the solutions in the cellular material. Time
periodicity is present in the frequency-domain nature of the reduced wave
equation, while space periodicity explicitly appears in the wave number
representation of the boundary conditions. The periodic relationship between
opposite sides of the fundamental cell, appearing in the boundary terms, allows
characterization of the fundamental properties of the material with the analysis
of a single cell. At the same time the wave vector \(\textbf{k}\) in
\eqref{eq:bvp_bloch} simultaneously describes: (i) the propagation direction of
a plane wave traveling through the unit cell and (ii) the spatial periodicity of
the plane wave. In consequence, finding solutions to the Bloch-BVP amounts to
finding those tuples \((\omega, \mathbf{k}, \mathbf{u})\) satisfying
\eqref{eq:bvp_bloch} when \(\mathbf{k}\) is varied in the dual Fourier based
representation of the fundamental material cell. This dual space corresponds to
the \textit{reciprocal space} and since it carries with it the periodic
character of the physical space it suffices to consider values (and directions)
of \(\mathbf{k}\) within this reciprocal space representation of the unit cell.

In the case of the C-CST medium, Bloch's theorem states that the eigenfunctions
of \eqref{eq:wave_eq_freq} can be expressed in the form
\[\mathbf{u}(\vb{x}) = \mathbf{u}(\vb{x}+\vb{a})e^{i\vb{k}\cdot\vb{a}} \]
where $\vb{a}$ is a vector that represents the periodicity of the material.
That is, the solution is the same at opposite sides of the unit cell, except 
for a phase shift factor $e^{i\vb{k}\cdot\vb{a}}$. Due to the linearity of the
differential equations we also have Bloch-periodic boundary 
conditions for the corresponding rotation and traction vectors.
Thus, in the case of the C-CST elastic solid, Bloch's theorem reduces to the
following set of boundary conditions for displacements, rotations,
force-tractions and couple-tractions in index notation:
\begin{subequations}
\begin{align}
    &u_i(\vb{x}) = u_i(\vb{x}+\vb{a})e^{i\vb{k}\cdot\vb{a}}\, , \\ 
    &\theta_i(\vb{x}) = \theta_i(\vb{x}+\vb{a})e^{i\vb{k}\cdot\vb{a}}\, ,\\
    &t_i(\vb{x}) = -t_i(\vb{x}+\vb{a})e^{i\vb{k}\cdot\vb{a}}\, , \label{eq:bloch_bcs.3}\\ 
    &m_i(\vb{x}) = -m_i(\vb{x}+\vb{a})e^{i\vb{k}\cdot\vb{a}}\, .
    \label{eq:bloch_bcs.4}
\end{align}
\label{eq:bloch_bcs}
\end{subequations}

\noindent The set of conditions summarized in \eqref{eq:bloch_bcs} will be
satisfied in a variational sense using a finite element formulation, where the
first two are essential boundary conditions and the other two natural boundary
conditions. Subsequently, a numerical model of the unit cell resulting in a
generalized eigenvalue problem will be solved for various specifications of the
wave vector.

\subsection{Hermiticity}
Our finite element algorithm follows from the action functional formulated in
\eqref{eq:Lagrangian_functional}. As discussed previously this amounts to the
solution of the weak form of the frequency domain reduced wave equations subject
to Bloch-periodic boundary conditions, as given by \eqref{eq:bloch_bcs}.
Neglecting body forces in \eqref{eq:Lagrangian_functional}, we have:
\begin{equation}
\begin{split}
\mathcal{A}[u; \omega] = &\frac{1}{2}\int\limits_{V} e_{ij}^* C_{ijkl} e_{ij} \dd{V}
  + \frac{1}{2}\int\limits_{V} \kappa_{i}^* D_{ij} \kappa_{i } \dd{V}
  -\frac{\omega^2}{2}\int\limits_{V} u_i^* \rho u_i \dd{V}\\
&- \int\limits_{S_t}  u_i^* t_i \dd{S}  - \int\limits_{S_m} \theta_i^* m_i \dd{S}\, ,
\end{split}
\label{eq:hamilton_freq_free}
\end{equation}

To obtain real eigenvalues that correspond to propagating waves in the band
structure of the material, the matrices resulting from the finite element
discretization must be Hermitic. Equivalently, we must prove  Hermiticity
(self-adjointness) in the action functional. This amounts to showing that the
boundary terms in \eqref{eq:hamilton_freq_free} vanish under Bloch periodic
boundary conditions.

Substitution of \eqref{eq:bloch_bcs} into surface integral terms of
\eqref{eq:hamilton_freq_free} yields
\begin{equation}
\begin{split}
&\int\limits_S u_i^{*}(\vb{x})t_i(\vb{x}) \dd{S} + \int\limits_S 
\theta_i^{*}(\vb{x}) m_i(\vb{x}) \dd{S} = \\
&\sum\limits_q \left\lbrace \int\limits_{S_q}\left[u_i^{*}(\vb{x})t_i(\vb{x}) + 
u_i^{*}(\vb{x} + \vb{a}_q)t_i(\vb{x} + \vb{a}_q)\right]\dd{S}_q + \right.\\
&\left. \int\limits_{S_q}\left[\theta_i^{*}(\vb{x}) m_i(\vb{x}) + 
\theta_i^{*}(\vb{x} + \vb{a}_q) m_i(\vb{x} + \vb{a}_q)\right]\dd{S}_q 
\right\rbrace \, ,
\end{split}
\end{equation}
with the index $q$ referring to each pair of opposite sides of the boundary.
Introducing the phase shifts and pulling out the common factors give:
\begin{equation}
\begin{split}
&\int\limits_S u_i^{*}(\vb{x})t_i(\vb{x}) \dd{S} + \int\limits_S 
\theta_i^{*}(\vb{x}) m_i(\vb{x}) \dd{S} = \\
&\sum\limits_q \left\lbrace \int\limits_{S_q} u_i^*(\vb{x})\left[t_i(\vb{x}) 
+ e^{i\vb{k}\cdot\vb{a}}t_i(\vb{x} + \vb{a}_q)\right]\dd{S}_q + \right.\\
&\left. \int\limits_{S_q}\theta_i^*(\vb{x})\left[m_i(\vb{x}) + 
e^{i\vb{k}\cdot\vb{a}}m_i(\vb{x} + \vb{a}_q)\right]\dd{S}_q \right\rbrace 
\, ,
\end{split}
\end{equation}
which after substituting 
\eqref{eq:bloch_bcs.3} and \eqref{eq:bloch_bcs.4}
leads to the vanishing of the boundary terms, thus proving the Hermiticity
condition.

\subsection{Positive definiteness}
Similarly, the proof for positive (semi)-definiteness reduces to showing that
the action functionals are related in such a way that:
\begin{equation}
\omega^2 = 
\frac{\mathcal{U}[u;\omega]}{\mathcal{\tilde T}[u;\omega]}  \geq 0 \, ,
\label{eq:definiteness}
\end{equation}
where
\[\mathcal{U}[u;\omega] = \frac{1}{2}\int\limits_{V} e_{ij}^* 
C_{ijkl}e_{kl}\dd{V} + \frac{1}{2}\int\limits_{V} \kappa_{i}^* D_{ij} \kappa_{j}\dd{V} 
\, \]
and
\[\mathcal{\tilde T}[u;\omega] = \frac{1}{2}\int\limits_V u_i^* \rho u_i \dd{V} \,  , \]
with the latter deriving directly from $\mathcal{T}[u;\omega]$.

Note that we have used the general representation \(C_{ijkl}\) and \(D_{ij}\) for
the constitutive tensors. The functional \(\mathcal{U}[u;\omega]\) is positive as
long as these constitutive tensors are positive definite, which holds true if
they satisfy
\begin{align*}
C_{ijkl}  e_{ij} e_{kl} \geq 0\quad \forall \ e_{mn}\, ,\\
D_{ij}  \kappa_{i} \kappa_{j} \geq 0\quad \forall \ \kappa_{m}\, .
\end{align*}

For isotropic materials, this implies the following constraints for the material
parameters:
\[\mu >0\, ,\quad   3\lambda + 2\mu > 0\, ,\quad \eta > 0\, .\]
On the other hand, the condition  \(u_i \neq 0\), requires \(\mathcal{\tilde T}\)
to be different from zero and thus the condition required by
\eqref{eq:definiteness}. In the case of rigid body motion, \(\mathcal{U}\)
could be zero implying that the form  is positive semi-definite, while the form
\(\mathcal{\tilde T}\) is positive definite.

\section{Finite element formulation}
In this section, we derive a consistent finite element formulation for periodic
couple stress elastodynamics, as an extension of those formulated by
\cite{darrall2014} for the corresponding quasistatic problem and by
\cite{guarin2020} for periodic micropolar Bloch analysis.  In particular,
the \(C^1\) displacement continuity requirement is avoided by using a Lagrange
multiplier approach. Other finite element solutions in C-CST include a penalty
method for isotropic elastostatics \citep{chakravarty2017}, Lagrange multipliers
for centrosymmetric anisotropic elastostatics \citep{pedgaonkar2021} and mixed
variable methods for isotropic elastodynamics \citep{deng2016, deng2017}.

\subsection{Lagrange multiplier reformulation}
Consider now a modification of the action given in
\eqref{eq:hamilton_freq_free} to include Lagrange multipliers \(\lambda_i\) that
enforce compatibility between the displacement field \(u_i\) and an assumed
independent rotation field \(\theta_i\).  Thus, the modified action becomes
\begin{equation}
\begin{split}
\mathcal{\hat A}[u; \omega] = &\frac{1}{2}\int\limits_{V} e_{ij}^* C_{ijkl} e_{ij} \dd{V}
  + \frac{1}{2}\int\limits_{V} \kappa_{i}^* D_{ij} \kappa_{i } \dd{V}
  -\frac{\omega^2}{2}\int\limits_{V} u_i^* \rho u_i \dd{V}\\
&- \int\limits_{S_t}  u_i^* t_i \dd{S}  - \int\limits_{S_m} \theta_i^* m_i \dd{S}\\
&+ \int\limits_{V}\lambda_i^* (\epsilon_{ijk} u_{k,j} - 2\theta_i )\dd{V}\, .
\end{split}
\label{eq:hamilton_freq_with_multipliers}
\end{equation}

For stationarity, we require
\[\var{\hat{\mathcal{A}}} = \pdv{\hat{\mathcal{A}}}{u_i} \var{u_i}
+ \pdv{\hat{\mathcal{A}}}{\theta_i}\var{\theta_i}
+ \pdv{\hat{\mathcal{A}}}{\lambda_i} \var{\lambda_i} = 0\, , \]
which is equivalent to
\begin{equation}
\begin{split}
&\int\limits_{V} \delta e_{ij}^* C_{ijkl} e_{ij} \dd{V}
  + \int\limits_{V} \delta \kappa_{i}^* D_{ij} \kappa_{i } \dd{V}
  -\omega^2 \int\limits_{V} \delta u_i^* \rho u_i \dd{V}\\
&\qquad- \int\limits_{S_t}  \delta u_i^* t_i \dd{S}  - \int\limits_{S_m} \delta \theta_i^* m_i \dd{S}\\
&\qquad+ \int\limits_{V}\delta \lambda_i^* (\epsilon_{ijk} u_{k,j} - 2\theta_i )\dd{V}\\
&\qquad+ \int\limits_{V} (\epsilon_{ijk} \delta u_{k,j}^* - 2\delta \theta_i^* ) \lambda_i \dd{V}\, .
\end{split}
\label{eq:weak_form_with_multipliers}
\end{equation}

\Cref{eq:weak_form_with_multipliers} is the modified weak form that will be used
here as the basis for the finite element Bloch analysis of an elastic
couple-stress solid. The Lagrange multiplier terms enforce the required
kinematic constraint between the continuum rotations \(\epsilon_{ijk} u_{k,j}\)
of the material point and the independent rotational variables \(\theta_i\).

From \eqref{eq:weak_form_with_multipliers}, we obtain the following
Euler-Lagrange equations
\begin{equation}
\begin{split}
&(C_{ijkl} e_{kl} + \epsilon_{ijk} \lambda_k)_{,j} = -\rho 
\omega^2u_i\quad \text{in } V,\\
&\epsilon_{ijk} (D_{kl} \kappa_{l})_{,j} - 2\lambda_i = 0\quad \text{in } V,\\
&\theta_i = \frac{1}{2}\epsilon_{ijk} u_{k,j}\quad \text{in } V,\\
&t_i = (C_{ijkl} e_{kl} + \epsilon_{ijk}\lambda_k)n_j\quad   \text{on } S_t,\\
&m_i = \epsilon_{ijk} D_{kl} \kappa_{l} n_j\quad   \text{on } S_m\, ,
\end{split}
\label{eq:bvp}
\end{equation}
Comparing this with \eqref{eq:skew_symmetry}, we can conclude that the Lagrange
multipliers equal the skew-symmetric part of the force-stress tensor, i.e.,
\[\lambda_i = s_i\, .\]

\subsection{Discretization}
\begin{figure}[H]
\centering
\includegraphics[width=2 in]{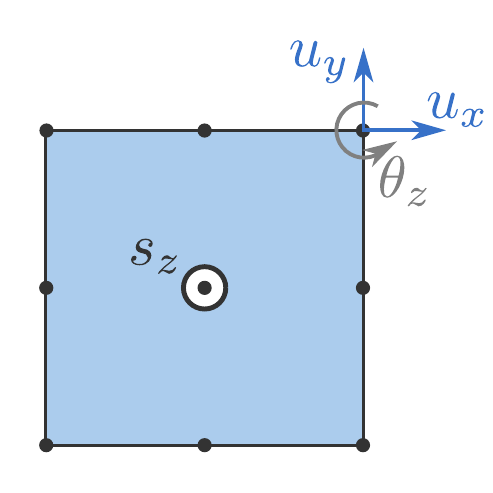}
\caption{Finite element used for the finite element discretization of the
C-CST material model. A second-order Lagrange interpolation is used for
displacements and rotations while a constant is used for the skew-symmetric
stresses. The nodal degrees of freedom are depicted as black disks while the
white disk describes the element skew-symmetric force-stress degree of freedom.}
\label{fig:element}
\end{figure}

To discretize \eqref{eq:weak_form_with_multipliers}, we use for the
element-based shape functions second-order Lagrange interpolation for the
displacements and rotations and constant skew-symmetric stresses. This
translates into \(C^0\) inter-element displacement and rotation continuity, and
skew-symmetric stresses that are constant within the element but discontinuous
between elements. \Cref{fig:element} depicts a typical element for the
discretization and the degrees of freedom used in two-dimensional idealizations.

To write the discretized equations, we will use a combined index notation. In
this context subscripts will still make reference to scalar components of
tensors while capital superscripts will indicate interpolation operations. For
instance in the expression
\[u_i = _{u}\hspace{-4pt}N_{i}^{Q}u^{Q}\]
subscripts $i$ indicate the scalar components of the vector \(u\). To facilitate
further operations this subscript is also placed in the shape function resulting
in terms like \(_{u}N_{i}^{Q}\) and where the term \(u^{Q}\)  represents the
nodal point displacement associated to the $Q$th nodal point. This nodal vector
implicitly considers horizontal and vertical rectangular components. To clarify,
the displacement interpolation scheme written here as
\(u_i = _{u}\hspace{-4pt}N_{i}^{Q}u^{Q}\) takes the following explicit form for
the single nodal point Q:
\begin{equation}
\begin{bmatrix}u_x\\ u_y\end{bmatrix}=
\left[\cdots\begin{array}{cc}N^Q &0\\ 0 &N^Q\end{array}\cdots\right]
\begin{Bmatrix}\vdots\\ u_x^Q\\ u_y^Q\\\vdots\end{Bmatrix} \, .
\label{eq:interp_nota}
\end{equation}

With this notation we write for the primary variables
\((u_i, \theta_i, s_i )\) the following interpolated versions
\begin{equation}
u_i = _{u}\hspace{-4pt}N_{i}^{Q}u^{Q},\quad \theta_i
  = _{\theta}\hspace{-4pt}N_i^{Q}\theta^{Q} ,\quad s_i
  = _{s}\hspace{-4pt}N_{i}^{Q} s^{Q} \, ,
\label{eq:interp_fun_prim}
\end{equation}
and similarly for the secondary kinematic descriptors
\(e_{ij} , \epsilon_{ijk} u_{i,j}\) and \(\kappa_{i}\) 
\begin{equation}
e_{ij} = _{e}\hspace{-4pt}B_{ij}^{Q}u^{Q} ,\quad \epsilon_{ijk} u_{i,j}
  = _{\nabla}\hspace{-4pt}B_{k}^{Q}u^{Q} ,\quad \kappa_{i}
  = _{\kappa}\hspace{-4pt}B_{i}^Q \theta^Q \, ,
\label{eq:interp_fun_second}
\end{equation}
together with the constitutive equations
\begin{equation}
\begin{aligned}
  &\sigma_{ij} = C_{ijkl}e_{kl}\, , \\
  &\mu_{i} = D_{ij}\kappa_{j}\, .
\end{aligned}
\label{eq:const}
\end{equation}

Substitution of the above relations in \eqref{eq:weak_form_with_multipliers}
gives the discrete version of the first variation of the modified correlated
action;
\begin{equation}
\begin{split}
&\var{\hat{\mathcal{A}}} = \delta u^{Q*} \int\limits_{V} (_{e}B_{ij}^{Q}) (C_{ijkl}) (_{e}B_{kl}^{P}) \dd{V} u^{P} - \rho \omega ^2 \delta u^{Q*}  \int\limits_{V} (_{u}N_{i}^{Q}) (_{u}N_{i}^{P}) \dd{V} u^{P} \\
&- \delta u^{Q*}\int\limits_{V}\ _{u}N_{i}^Q f_i \dd{V} - \delta u^{Q*}\int\limits_{S}\ _{u}N_{i}^Q t_i \dd{S} 
+ \delta \theta^{Q*} \int\limits_{V} (_{\kappa}B_{i}^{Q}) (D_{ij}) (_{\kappa}B_{j}^{P}) \dd{V} \theta^{P}  \\
&-\delta \theta^{Q*}\int\limits_{S}\ _{\theta}N_i^Q m_i \dd{S} + \delta s^{Q*} \int\limits_{V} (_{s}N_{k}^{Q}) (_{\nabla}B_{k}^{P}) \dd{V}  u^{P}
+ \delta u^{Q*} \int\limits_{V} (_{\nabla}B_{k}^{Q}) (_{s}N_{k}^{P})  \dd{V} s^{P} \\
&-\delta s^{Q*} \int\limits_{V} 2 (_{s}N_{k}^{Q}) (_{\theta}N_{k}^{P}) \dd{V} \theta^{P} -\delta \theta^{Q*} \int\limits_{V} 2(_{\theta}N_{k}^{Q}) (_{s}N_{k}^{P}) \dd{V} s^{P} = 0 \, .
\end{split}
\label{eq:discrete_PVW}
\end{equation}

The explicit form of the interpolators defined above is given in the appendix.

\subsection{Discrete equilibrium equations}
From the arbitrariness in the variations \(\delta u^Q\) , \(\delta \theta^Q\)
and \(\delta s^Q\) in \eqref{eq:discrete_PVW} it follows that:
\begin{equation*}
\begin{split}
&\int\limits_{V} (_{e}B_{ij}^{Q}) (C_{ijkl}) (_{e}B_{kl}^{P}) \dd{V} u^P
  - \int\limits_{V} (_{u}N_{i}^{Q}) (_{u}N_{i}^{P}) \dd{V} u^P
  - \int\limits_{V}\ _{u}N_{i}^Q f_i \dd{V}
  - \int\limits_{S}\ _{u}N_{i}^Q t_i \dd{S} = 0\,, \\
&\int\limits_{V} (_{\kappa}B_{i}^{Q}) (D_{ij}) (_{\kappa}B_{j}^{P}) \dd{V} \theta^P 
  - \int\limits_{S}\ _{\theta}N_i^Q m_i \dd{S}
  - \int\limits_{V} 2(_{\theta}N_{k}^{Q}) (_{s}N_{k}^{P}) \dd{V} s^P=0\, , \\
&\int\limits_{V} (_{s}N_{k}^{Q}) (_{\nabla}B_{k}^{P}) \dd{V}  u^P
  - \int\limits_{V} 2 (_{s}N_{k}^{Q}) (_{\theta}N_{k}^{P}) \dd{V} \theta^P = 0\, ,
\end{split}
\end{equation*}
which can be written in the standard finite element form for dynamic equilibrium
\begin{equation}
\begin{bmatrix}
K_{uu}^{QP} &0 &K_{us}^{QP}\\
0 &K_{\theta\theta}^{QP} &-K_{\theta s}^{QP}\\
K_{s u}^{QP} &-K_{s\theta}^{QP} &0
\end{bmatrix}
\begin{Bmatrix}
u^P\\
\theta^P\\
s^P
\end{Bmatrix}
=
\omega^2\begin{bmatrix}
 M_{uu}^{QP} &0 &0\\
0 &0 &0\\
0 &0 &0
\end{bmatrix}
\begin{Bmatrix}
u^P\\
\theta^P\\
s^P
\end{Bmatrix}
+ \begin{Bmatrix}
F_u^Q\\
m_\theta^Q\,\\
0\end{Bmatrix}
\label{eq:mat_fem}
\end{equation}
where the individual terms are defined as
\begin{align*}
K_{uu}^{QP} &= \int\limits_{V} (_{e}B_{ij}^{Q}) (C_{ijkl}) (_{e}B_{kl}^{P}) \dd{V}\, ,
&M_{uu}^{QP} &= \rho \omega ^2 \int\limits_{V} (_{u}N_{i}^{Q}) (_{u}N_{i}^{p}) \dd{V}\, ,\\
K_{u s}^{QP} &= \int\limits_{V} (_{\nabla}B_{k}^{Q}) (_{s}N_{k}^{P})  \dd{V}\, ,
&F_{u }^{Q} &= \int\limits_{V}\ _{u}N_{i}^Q f_i \dd{V} + \int\limits_{S}\ _{u}N_{i}^Q t_i \dd{S}\, ,\\
K_{\theta \theta}^{QP} &= \int\limits_{V} (_{\kappa}B_{i}^{Q}) (D_{ij}) (_{\kappa}B_{j}^{P}) \dd{V}\, ,
&K_{\theta s}^{QP} &= \int\limits_{V} 2(_{\theta}N_{k}^{Q}) (_{s}N_{k}^{P}) \dd{V}\, ,\\
m_{\theta}^{Q} &= \int\limits_{S}\ _{\theta}N_i^Q m_i \dd{S}\, ,
&K_{s u}^{QP} &= \int\limits_{V} (_{s}N_{k}^{Q}) (_{\nabla}B_{k}^{P}) \dd{V}\, ,\\
K_{s \theta}^{QP} &= \int\limits_{V} 2 (_{s}N_{k}^{Q}) (_{\theta}N_{k}^{P}) \dd{V}\, .
\end{align*}

\Cref{eq:mat_fem} can be rewritten in the following set of equilibrium equations
in terms of nodal forces and couples
\begin{equation}
\begin{aligned}
f_{(\sigma)}^Q + f_{s}^Q - f_I^Q - T^Q &= 0\, ,\\
m_{\mu}^Q + m_{s}^Q - q^Q &= 0\, ,\\
s(\theta - \hat{\theta}) = 0\, ,
\end{aligned}
\label{eq:discrete_balance}
\end{equation}
where the subindex $(\sigma)$ refers to the symmetric part of the stress tensor,
and \(I\) to inertial forces. Notice that we do not have an
inertial term for the second equation as is the case for the micropolar model
\citep{guarin2020}. We also have a third equation reflecting the kinematic
restriction, between the rotation \(\theta\) and the introduced degree of
freedom \(\hat{\theta}\),  imposed via the Lagrange-multiplier term \(s\) in
each element.

When using a Lagrange multiplier formulation as in
\eqref{eq:weak_form_with_multipliers} the equations are still self-adjoint, as
can be seen in the structure of \eqref{eq:mat_fem}. Nevertheless, the stiffness
matrix is indefinite and the solution of the problem represents a saddle-point
instead of a minimum \citep{arnold_mixed_1990, darrall2014}.

\subsection{Eigenvalue problem}
In finding the dispersion relations, we are interested in the free wave motion
in the media. This leads to the following eigenvalue problem
\begin{equation}
[K]\{U\} = \omega^2 [M]\{U\}
\label{eq:eigenvalue_fem}
\end{equation}
with
\[[K] = \begin{bmatrix}
K_{uu}^{QP} &0 &K_{us}^{QP}\\
0 &K_{\theta\theta}^{QP} &-K_{\theta s}^{QP}\\
K_{s u}^{QP} &-K_{s\theta}^{QP} &0
\end{bmatrix} ,\,
[M] = \begin{bmatrix}
M_{uu}^{QP} &0 &0\\
0 &0 &0\\
0 &0 &0
\end{bmatrix}, \,
\{U\} = 
\begin{Bmatrix}
u^P\\
\theta^P\\
s^P
\end{Bmatrix}\, .\]

In \eqref{eq:eigenvalue_fem} Bloch-periodic boundary  conditions are yet to be
imposed. This can be done in two ways \citep{valencia_uel_2019}: (i) modifying
the connectivity of the elements; and (ii) assembling the matrices without
considering boundary conditions and impose the Bloch-periodicity through
row/column operations. In this work, we follow the second approach as it
requires the stiffness and mass matrices to be assembled once and the
transformation matrices are computed for every wavenumber in the first
Brillouin zone. This process results in the following eigenvalue problem
\begin{equation}
[K_R(\vb{k})]\{U\} = \omega^2 [M_R(\vb{k})]\{U\}
\label{eq:reduced_eigenvalue_fem}
\end{equation}
with
\[[K_R(\vb{k})] = [T(\vb{k})^H K T(\vb{k})]\, ,\quad [M_R(\vb{k})] = [T(\vb{k})^H M T(\vb{k})]\, ,\]
where \([T(\vb{k})]\) represents the transformation matrix for a given \(\vb{k}\),
and the \([T^H]\) refers to the Hermitian transpose of \([T]\). For an explicit
form for the matrices \([T]\) refer to \cite{hussein2014dynamics} or
\cite{guarin2012_msc}.

We conducted the implementation on top of the in-house finite element code
SolidsPy \citep{solidspy} and used SciPy to solve the eigenvalue problem
\citep{scipy}. To take advantage of the sparsity of the matrices the problem
should be written as matrix-vector multiplications, such as
\begin{align*}
\{x\} = [T]\{U\}\, ,\\
\{y\} = [K]\{x\}\, ,\\
\{z\} = [T^H] \{y\}\, ,
\end{align*}
with $\{z\}$ representing the image of the linear operator \([K_R]\) over
\(\{U\}\). The same procedure can be applied for the right-hand side of
\eqref{eq:reduced_eigenvalue_fem}.

The Lagrange-multiplier approach represents a saddle-point instead of a minimization
problem \citep{arnold_mixed_1990}. This can be seen in the structure of the stiffness
matrix obtained in equation \eqref{eq:eigenvalue_fem}. Furthermore the mass matrix
is not positive definite anymore. This structure for the eigenvalue problem requires
the use of a specific solver such as the LOBPCG method \citep{knyazev2001}
instead of the classical Arnoldi method \citep{book:arpack}.

\section{Results: Dispersion relations for C-CST cellular materials}
In this section we conduct a series of dispersion analyses intended to show the
effectiveness of our mixed finite element implementation of the C-CST material
model in predicting the correct wave propagation properties of the material. All
the dispersion graphs use the dimensionless frequency
\begin{equation}
\Omega = \frac{2d\omega}{c_2}\, ,
\end{equation}
for the vertical axis, where \(2d\) is the dimension of the unit cell and
\(c_2^2 = \mu/\rho\) is the speed of the shear wave for a classical elastic material.
The Poisson ratio for all the simulations is \(\nu = 1/4\).

As a first instance we find the response of a homogeneous periodic material
which has also a closed form solution. We will then continue to study a second
prototypical example corresponding to a homogeneous material with a circular
pore. These two problems exhibit two different levels of dispersive behavior. In
the homogeneous material cell, dispersion is due to the kinematic enrichment of
the model associated to the length scale parameter, while in the porous material
model additional dispersion arises due to the explicit microstructural feature.

\subsection{Homogeneous material}
\begin{figure}[H]
\centering
\includegraphics[height=2.8 in]{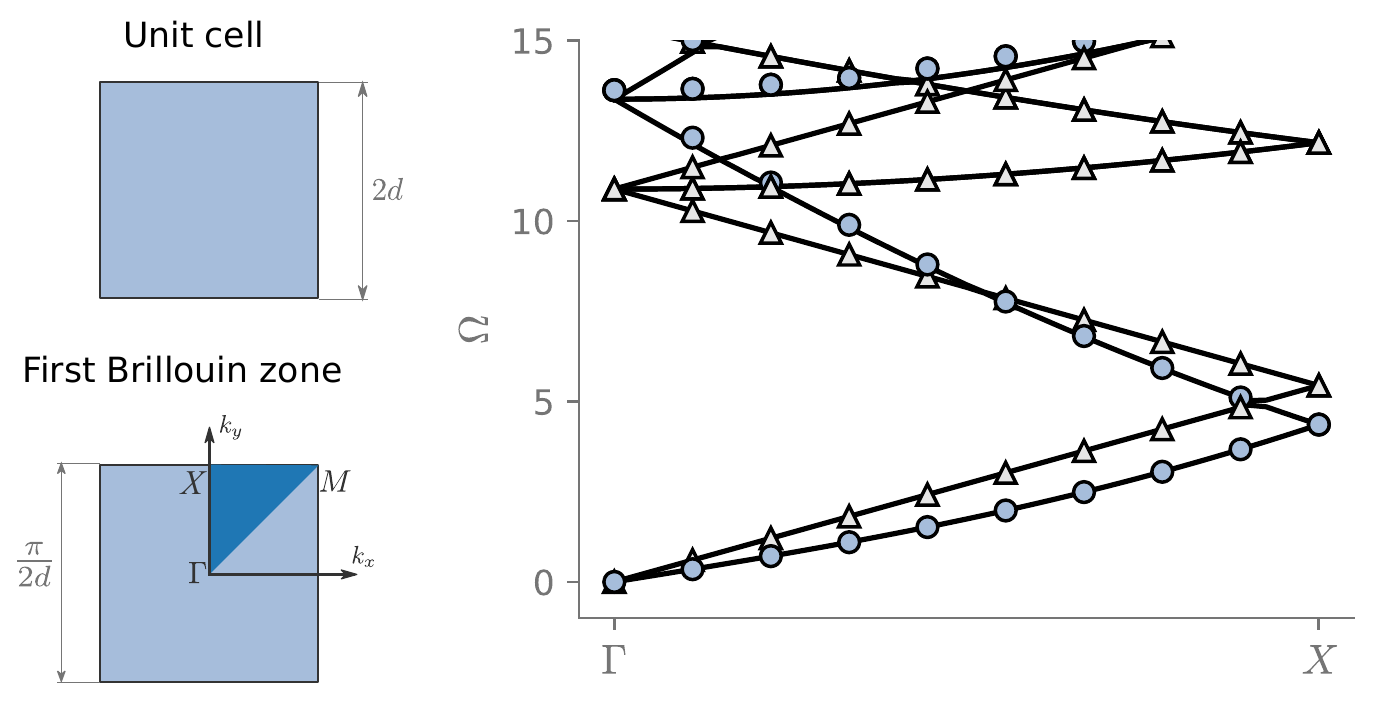}
\caption{Dispersion relations for a homogeneous material model. Solid lines
represent FEM results while markers correspond to the analytic solution.
Triangular and filled-dots describe the P and SV wave modes, respectively.}
\label{fig:homogeneous}
\end{figure}

As a test of accuracy and effectiveness of our implementation we consider the
case of a homogeneous material cell with the same mechanical properties of the
material reported previously and with closed form dispersion relations from
\eqref{eq:dispersion} and \eqref{eq:dispersion_SV}. In this model microstructural
effects are introduced through the material length parameter \(\ell\).
Recall that \(\ell^2\) is defined by the ratio \(\frac{\eta}{\mu}\) where
\(\eta\) is the curvature-couple-stress module while \(\mu\) is the shear
modulus from Cauchy elasticity. The results in terms of the resulting band
structure are shown in \cref{fig:homogeneous}, where we used a
\(16\times16\) mesh and \(\ell^2/d^2 = 3/8\). For a conceptual description of
the reciprocal space and a guide on how to interpret the results in a Bloch
analysis the reader is referred to \cite{Valencia_periodic}. Note that this set
of results is directly comparable with the curves from the closed form solutions
from \cref{fig:dispersion_analytic}. Since the material is isotropic there are
no directional effects and, as discussed previously, the only difference between
this model and the result from classical elasticity is the dispersive behavior
of the shear wave. In contrast with the micropolar model \citep{guarin2020},
the present C-CST model does not exhibit additional rotational waves.

\begin{figure}[H]
\centering
\includegraphics[width=5.5 in]{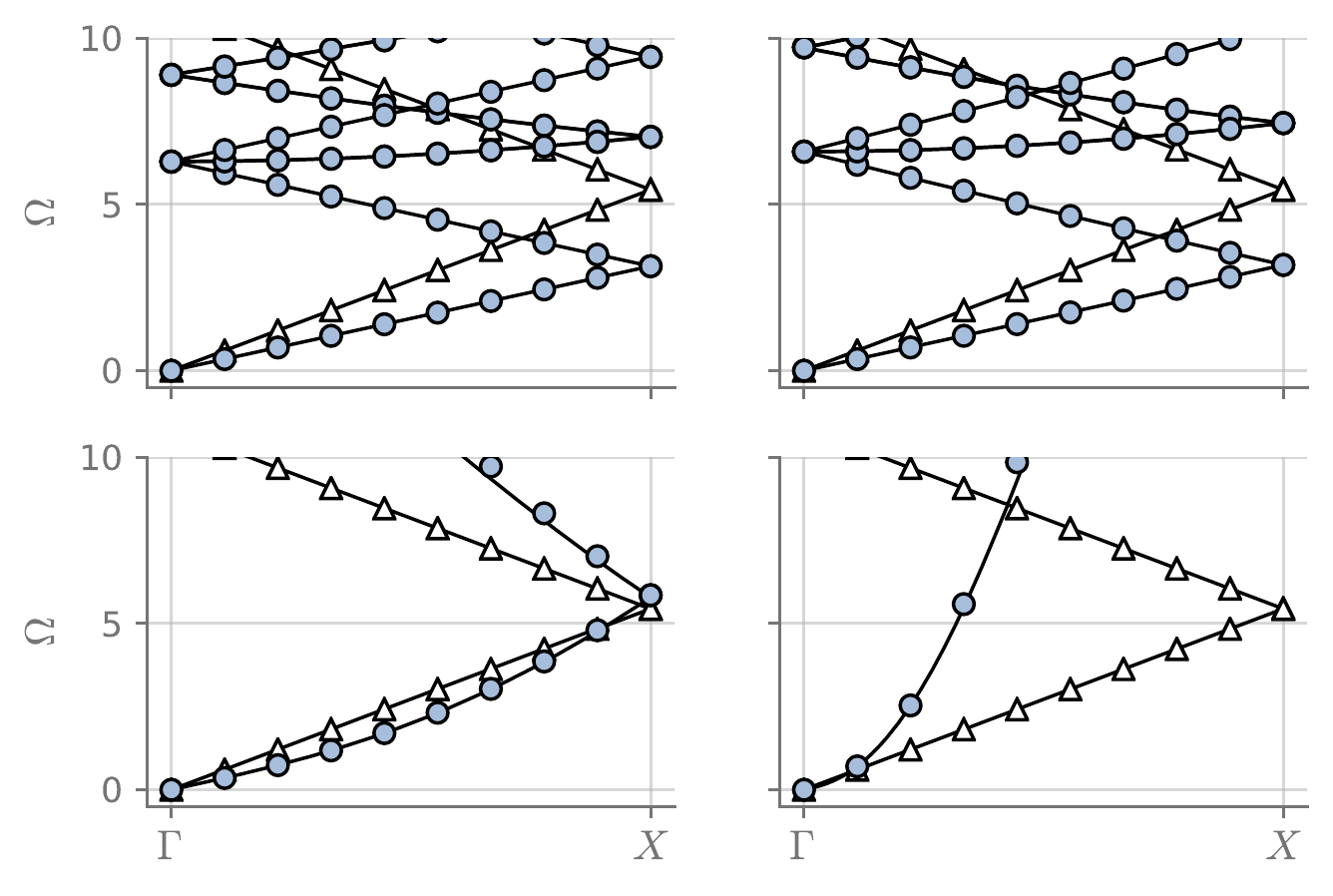}
\caption{Dispersion relations for a homogeneous material model with varying
length scales, \(\ell/d \in [0.01, 0.1, 1, 10]\). Solid lines represent FEM
results while markers correspond to analytic solutions. As expected, for
increasing \(\ell/d\) the S-wave presents more dispersion due to the factor
\(\sqrt{1 + k^2 \ell^2}\), as presented in \eqref{eq:dispersion_SV}.}
\label{fig:homog_lengths}
\end{figure}

\Cref{fig:homog_lengths} shows the results for the same
material cell but now we have considered 4 different values of the length scale
parameter corresponding to \(\ell/d \in [0.01, 0.1, 1, 10]\). The mesh in this
case is \(16\times16\). Notice that, as expected, the increasing value of this
parameter only affects the dispersive response of the shear waves while the
P-waves retain their classical non-dispersive behavior. As seen in
\eqref{eq:dispersion_SV} the dispersion increases for higher values of \(\ell\)
due to the factor \(\sqrt{1 + k^2 \ell^2}\) in the dispersion relation. This
behavior is closely followed by the numerical results presented in
\cref{fig:homog_lengths}.

\begin{figure}[H]
\centering
\includegraphics[width=5.75 in]{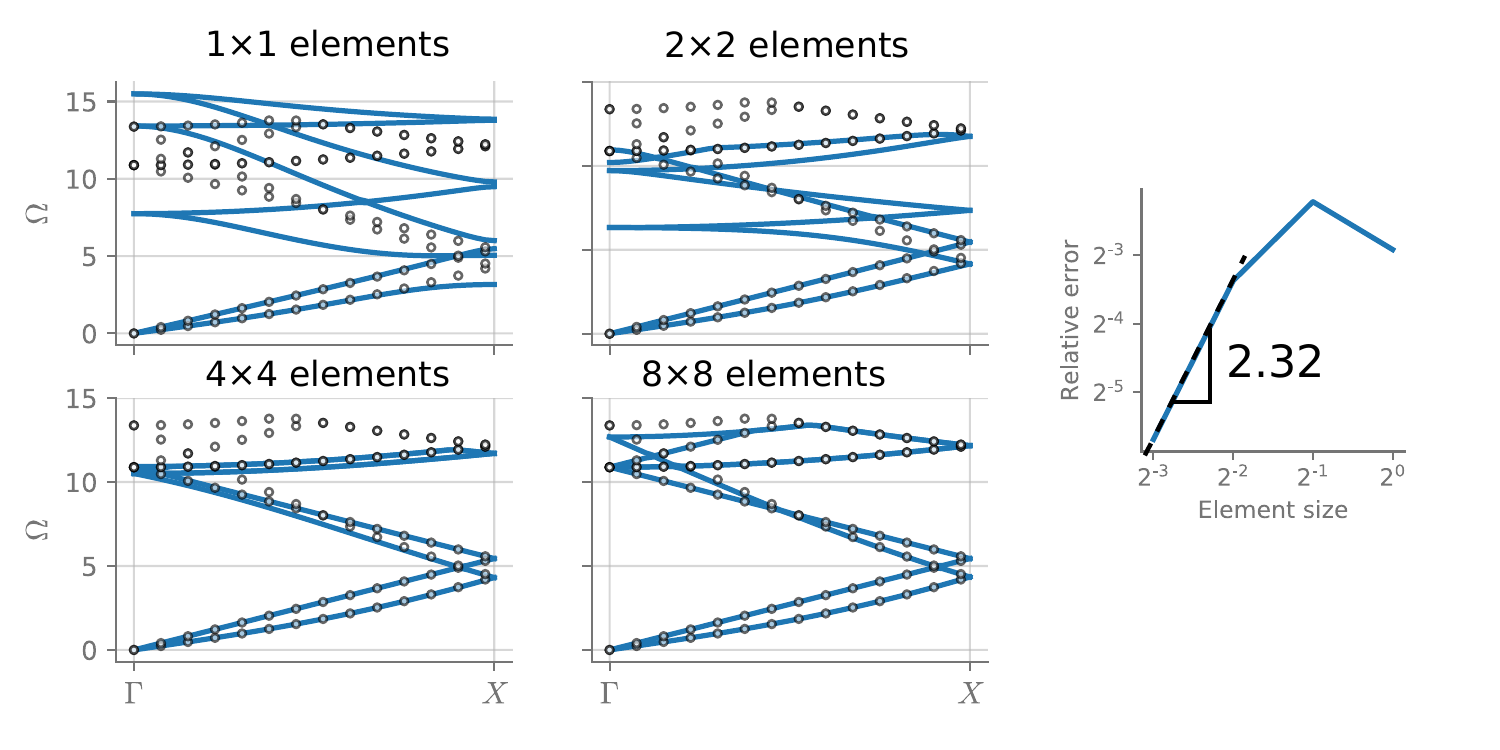}
\caption{Convergence of the first 8 modes in the dispersion relations at
\(\ell^2/d^2 = 3/8\) for a sequence of meshes with: \(1\times1\), \(2\times2\),
\(4\times4\), and \(8\times8\) elements --- presented as solid blue lines in the 
background. The results are compared with a mesh that has \(16\times16\)
elements --- presented as dots in the foreground. The estimated convergence rate
for the eigenvalues in the 2-norm is 2.32.}
\label{fig:convergence}
\end{figure}

As an additional verification, we also tested the convergence
in the  calculation of the dispersion relations after considering the first 8
modes  for a sequence of meshes of \(1\times1\),  \(2\times2\), \(4\times4\),
and \(8\times8\) elements for \(\ell^2/d^2 = 3/8\). The error in the eigenvalue
computation was measured according to
\[e = \frac{\Vert\vb*{\omega}_\text{ref} - 
\vb*{\omega}_h\Vert_2}{\Vert\vb*{\omega}_\text{ref}\Vert_2}\, ,\]
where \(\vb*{\omega}_h\) is the set of eigenvalues (dispersion relation) for a 
mesh of characteristic element size \(h\) and \(\vb*{\omega}_\text{ref}\) is the 
solution corresponding to the finer \(16\times16\) elements mesh, which has been
taken as reference. The results for this sequence, together with the variation
in the  error parameter, are displayed in \cref{fig:convergence}. The estimated 
convergence rate for the eigenvalues is 2.32. We see that when
we refine the mesh it can reproduce the dispersion curves better for higher
frequencies. There are still some differences between the \(8\times8\) and
\(16\times16\) meshes around the dimensionless frequency of 15 but these
differences will disappear with further refinement. Nevertheless, opposed to
what happens in classical continua we would need more points per wavelength
every time that we want to increase the maximum frequency. This is due to the
inherent dispersive behavior of S-waves as can be seen in equation
\eqref{eq:dispersion_SV}. Thus, we would expect to need more than 10 points per
wavelength, customary for finite element methods, or 5, customary for spectral
element methods \citep{komatitsch1999introduction, ainsworth2009dispersive,
guarin2015}. Again, we should emphasize the dispersive nature of the SV-waves,
while the P-waves remain non-dispersive.

\subsection{Dispersion in cellular material with a circular pore}
Now, we consider a C-CST composite material cell configured by a circular pore
embedded in a homogeneous matrix. The presence of the pore provides the model
with a second length-scale due to the microstructure, in addition to the one
introduced by the material length-scale parameter \(\ell\). For
illustration, we assume a pore diameter \(a\) that is half the cell length
(i.e., \(a=d\)). Thi is equivalent to a porosity of \(\pi/16\) or approximately
0.196, which is kept fixed as we modify the size of the unit cell to control
\(\ell/a\).

\begin{figure}[H]
\centering
\includegraphics[width=5.5 in]{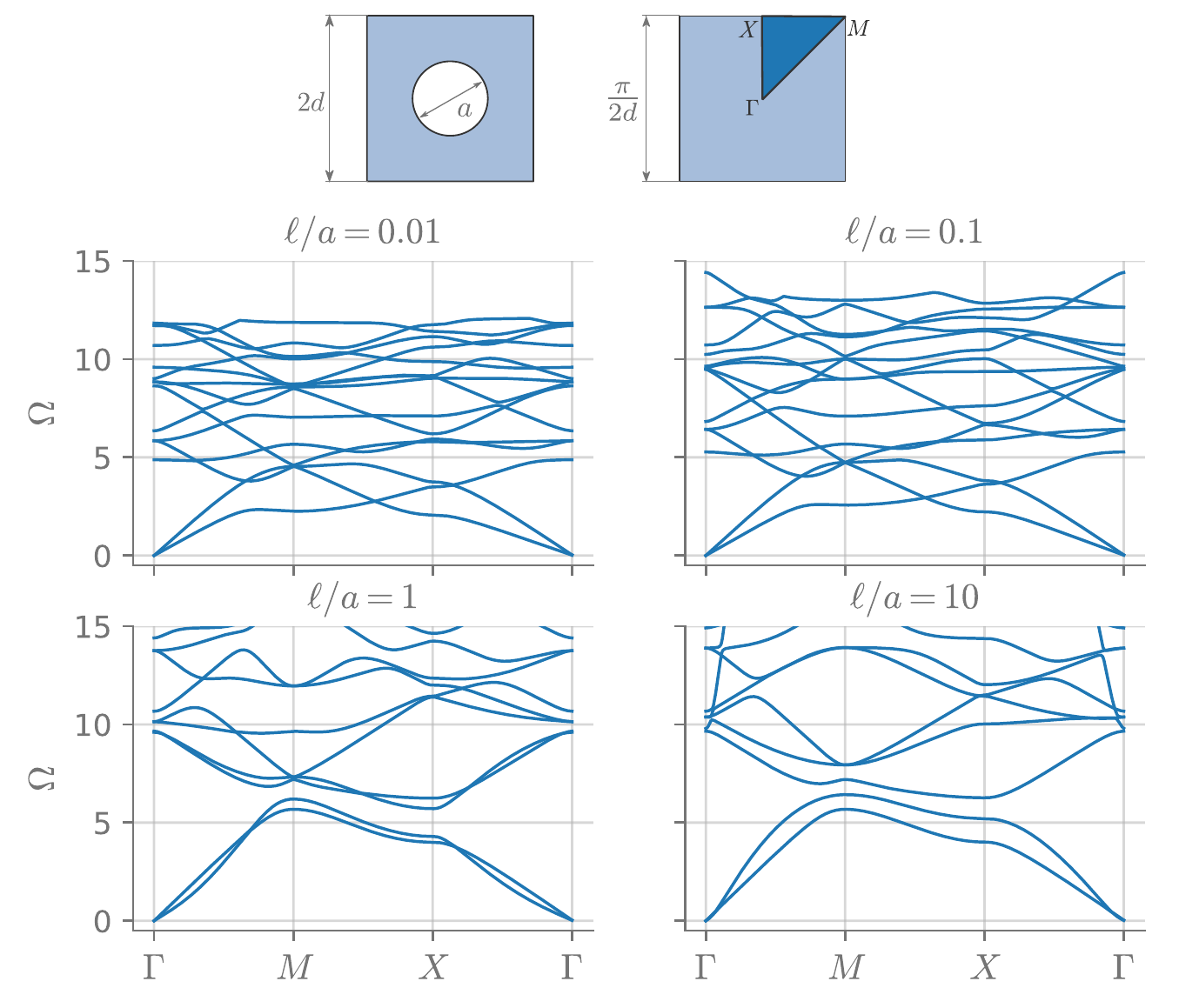}
\caption{Dispersion for a cellular material with circular pores for varying 
length scales, $\ell/a = [0.01, 0.1, 1, 10]$.}
\label{fig:pore}
\end{figure}

The resulting dispersion curves for this cellular material
with four different length scale ratios are shown in \cref{fig:pore}
with \(\ell/a = [0.01, 0.1, 1, 10]\). In contrast to the results from the fully
homogeneous material cell, the presence of the circular pore introduces
scattering effects inside each cell and the composite shows much more
complicated elastodynamic behavior.  Most importantly, however, the dispersion
curves become more regular with increased \(\ell/a\) and partial bandgaps
open up along the \(\Gamma M\) and \(X\Gamma\) directions, especially for
\(\ell/a = 1\) and \(\ell/a = 10\). This type of band structure is not
observed for classical elastodynamic cells with a similar geometric periodicity,
which would exhibit behavior close to that obtained here with \(\ell/a = 0.01\).
In fact, as \(\ell/a \rightarrow 0\), C-CST theory recovers the classical
result.

Furthermore, it is important to note that this interesting band structure is
obtained with a consistent continuum mechanics formulation, which requires only
a single additional material parameter, \(\ell\), beyond those needed in the
classical elastic case. When this intrinsic length scale is on the order of the hole
diameter, the dispersive SV wave has a fundamental group velocity for the cell
with dimension \(2d\) approximately equal to the group (and phase) velocity of
the non-dispersive P wave, which allows the SV and P branches to follow a
similar path, causing band gaps to open. This behavior, which is not seen in
classical elastodynamics, occurs in the regions near to where the two branches
intersect in \cref{fig:dispersion_analytic}. Consequently, under C-CST,
these band gaps will originate whenever the size of the cellular structure is
tuned to the material length scale, a potentially significant phenomenon that
has not been recognized previously. With further tuning of the porosity level
and cell size, it may be possible to achieve even a complete bandgap at
relatively low non-dimensional frequency \(\Omega\).

\section{Conclusions}

The present work incorporates several innovative aspects. First of all, we
have developed a novel frequency domain correlated action principle for the
consistent couple stress theory (C-CST) of \cite{hadjesfandiari2011couple} and
used that to extend the Lagrange multiplier finite element algorithm of
\cite{darrall2014} to study periodic cellular materials through Floquet-Bloch
theory from solid state physics. Particularly, we have addressed the imposition
of extended Bloch boundary conditions for this material model where in
addition to force tractions and displacements there are also couple tractions
and rotations. Secondly, we also discussed numerical aspects related to the
solution of the wavenumber dependent generalized eigenvalue problem resulting
from the imposition of the Bloch periodic boundary conditions, overcoming
complications arising from the inclusion of Lagrange multipliers and a
non-positive definite mass matrix. The implementation was
shown to give accurate results for homogeneous and porous unit cells and for
varying couple stress material length-scale parameters.

The analysis of the first cell was used mainly to test the correctness of our
implementation as this material has a closed-form dispersion relation. The
algorithm was shown to correctly capture the non-dispersive P-wave as well as
the dispersive SV-wave. This analysis was complemented by a convergence analysis
with four different meshes of increasing refinement for the material cell. The
observed convergence rate shows that the Lagrange multiplier algorithm is
effective in maintaining continuity by imposing the newly introduced kinematic
constraint implicit in the mean curvature tensor definition. As the final
contribution, we have discovered the interesting bandgap structure of a material
cell with a circular pore embedded in a homogeneous matrix, which reveals the
appearance of bandgaps introduced by the kinematic features of C-CST and the
dispersive behavior of the SV-waves defined in terms of the microstructural
length scale parameter.

From a general perspective, C-CST is a true size-dependent continuum theory,
which is intended here for periodic elastic material cells at scales for which a
continuum representation is appropriate.  From the results shown in the paper,
C-CST becomes important when the size of the cell in on the order of the
intrinsic length scale parameter or smaller.  For larger cells, the classical
theory can be used instead. On the other hand, micropolar theory disconnects the
rotational field from the displacements, which can lead to approximations that
may or may not be physical.

\appendix
\section{Explicit form of the finite element interpolators in the C-CST solid}
\label{app:dir}

In the case of isotropic materials under plane strain idealizations, we have
the following equations
\begin{align*}
&\left(\lambda + 2 \mu\right) \left(\frac{d^2 u_x}{d x^2}  + \frac{d^2 u_y}{d yd x} \right)
- \mu \left(\frac{d^2 u_y}{d yd x}  - \frac{d^2 u_x}{d y^2} \right)  
- \eta \left(\frac{d^4 u_x}{d y^4}  + \frac{d^4 u_x}{d y^2 d x^2} 
- \frac{d^4 u_y}{d y^3 d x} - \frac{d^4 u_y}{d y d x^3} \right) = -\rho\omega^2 u_x\\
&\left(\lambda + 2 \mu\right) \left(\frac{d^2 u_y}{d y^2}  + \frac{d^2 u_x}{d yd x} \right)
- \mu \left(\frac{d^2 u_x}{d y d x} - \frac{d^2 u_y}{d x^2}\right)
- \eta \left(\frac{d^4 u_y}{d x^4}  - \frac{d^4 u_x}{d y^3 d x} 
- \frac{d^4 u_x}{d y d x^3}  + \frac{d^4 u_y}{d y^2 d x^2} \right) = -\rho\omega^2 u_y
\end{align*}

We have the following explicit forms for the interpolation matrices in two
dimensions (\cite{book:bathe}): 
\begin{align*}
_{u}\mathbf{N}^Q &= _{\theta}\mathbf{N}^Q = _{s}\hspace{-2pt}\mathbf{N}^Q =
\begin{bmatrix}
\cdots &N^Q  &  0 &\cdots\\
\cdots & 0 &N^Q &\cdots
\end{bmatrix}\, ,
&_{e}\mathbf{B} = \begin{bmatrix}
\cdots & \frac{\partial N^Q}{\partial x}  & 0 &\cdots\\
\cdots & 0 &\frac{\partial N^Q}{\partial x}  &\cdots\\
\cdots & \frac{\partial N^Q}{\partial y}  & \frac{\partial N^Q}{\partial 
x}&\cdots
\end{bmatrix}\, ,\\
_{\kappa}\mathbf{B} &= \begin{bmatrix}
\cdots & -\frac{\partial N^Q}{\partial y}  &\cdots\\
\cdots & \frac{\partial N^Q}{\partial x}  &\cdots
\end{bmatrix}\, ,
&\mathbf{_{\nabla}B} = \begin{bmatrix}
\cdots & -\frac{\partial N^Q}{\partial y}  &\frac{\partial N^Q}{\partial x} 
&\cdots
\end{bmatrix}\, .
\end{align*}
an the following constitutive tensors in Voigt notation
\begin{align*}
&\mathbf{C} = \frac{E (1 - \nu)}{(1 + \nu) (1 - 2\nu)}\begin{bmatrix}
1 & \frac{\nu}{1 - \nu} &0\\
\frac{\nu}{1 - \nu} & 1 &0\\
0 & 0 &\frac{1 - 2\nu}{2(1 - \nu)}
\end{bmatrix}\, ,\quad 
&\mathbf{D} = 4\eta \begin{bmatrix}
1 & 0\\
0 &1
\end{bmatrix}\, .
\end{align*}

\section*{Acknowledgements}
This work was supported by Universidad EAFIT.

\bibliographystyle{plainnat}
\bibliography{bloch-cosserat}

\end{document}